\documentclass[aps,prl,twocolumn,showpacs,showkeys,footinbib,longbibliography,groupedaddress]{revtex4-1}

\usepackage{amsmath}
\usepackage{amssymb}
\usepackage{graphicx}
\usepackage{bm}
\usepackage{color}
\usepackage{relsize}
\usepackage{braket}
\usepackage{bbold}
\usepackage{txfonts}
\usepackage{epstopdf}
\usepackage{hyperref}

\begin{document}

\title{Tunable magnetic textures in spin valves: From spintronics to Majorana bound states}
\author{Tong Zhou$^1$}\
\email{tzhou8@buffalo.edu}
\author{Narayan Mohanta$^{2}$}
\author{Jong E. Han$^1$}
\author{Alex Matos-Abiague$^2$}
\author{Igor \v{Z}uti\'c$^{1}$}
\address{$^1$Department of Physics, University at Buffalo, State University of New York, Buffalo, New York 14260, USA\\
$^2$Department of Physics and Astronomy, Wayne State University, Detroit, MI 48201, USA}

\date{\today}

\begin{abstract}
Spin-valve structures in which a change of magnetic configuration is responsible for magnetoresistance led to 
impressive advances in spintronics, focusing on magnetically storing and sensing information. However, this mature technology also offers versatile control of magnetic textures with usually neglected underlying fringing fields to enable entirely different applications by realizing topologically-nontrivial states. Together with proximity-induced superconductivity in a two-dimensional electron gas with a large $g$-factor, 
these fringing fields realized in commercially-available spin valves provide Zeeman splitting, synthetic spin-orbit coupling, and confinement, needed for Majorana bound states (MBS). Detailed support for the existence and control of MBS is obtained by combining accurate micromagnetic simulation of fringing fields used as an input in Bogoliubov de Gennes equation to calculate low-energy spectrum, wavefunction localization, and local charge neutrality. A generalized condition for a quantum phase transition in these structures provides valuable guidance for the MBS evolution and implementing reconfigurable effective topological wires.
\end{abstract}

\maketitle

With non-Abelian statistics and nonlocal degrees of freedom, Majorana bound states (MBS) provide intriguing opportunities to implement topological quantum computing~\cite{Kitaev2001:PU,Nayak2008:RMP,Aasen2016:PRX,Elliott2015:RMP}. While there are impressive experimental advances in realizing 
MBS~\cite{Zhang2018:N,Albrecht2016:N,Deng2016:S,Mourik2012:S,Deng2012:NL,Rokhinson2012:NP} they rely on signatures, such as the quantized zero bias conductance peak~\cite{Sengupta2001:PRB,Law2009:PRL,Liu2017:PRB}, which do not verify the non-Abelian character. Some of the common 
obstacles in directly probing non-Abelian statistics through braiding or fusing MBS are inherent to 
one-dimensional (1D) geometries and thus it would be desirable to seek alternative platforms. One of them 
employs the interplay between the superconducting and magnetic proximity effects to generate topological states in a two-dimensional electron gas 
(2DEG)~\cite{Fatin2016:PRL,Matos-Abiague2017:SSC}. 
It is further stimulated by the demonstrated robust 
proximity-induced superconductivity in all-epitaxial 2D structures and a versatile control of spin valves~\cite{Shabani2016:PRB,Mayer2018:P,Wickramasinghe2018:APL,%
Sestoft2018:P,Ren2018:P,Fornieri2018:P,Kent2015:NN,Nowak2016:IEEEML}.

\begin{figure}[b]
\vspace{-0.55cm}
\includegraphics*[width=0.42\textwidth]{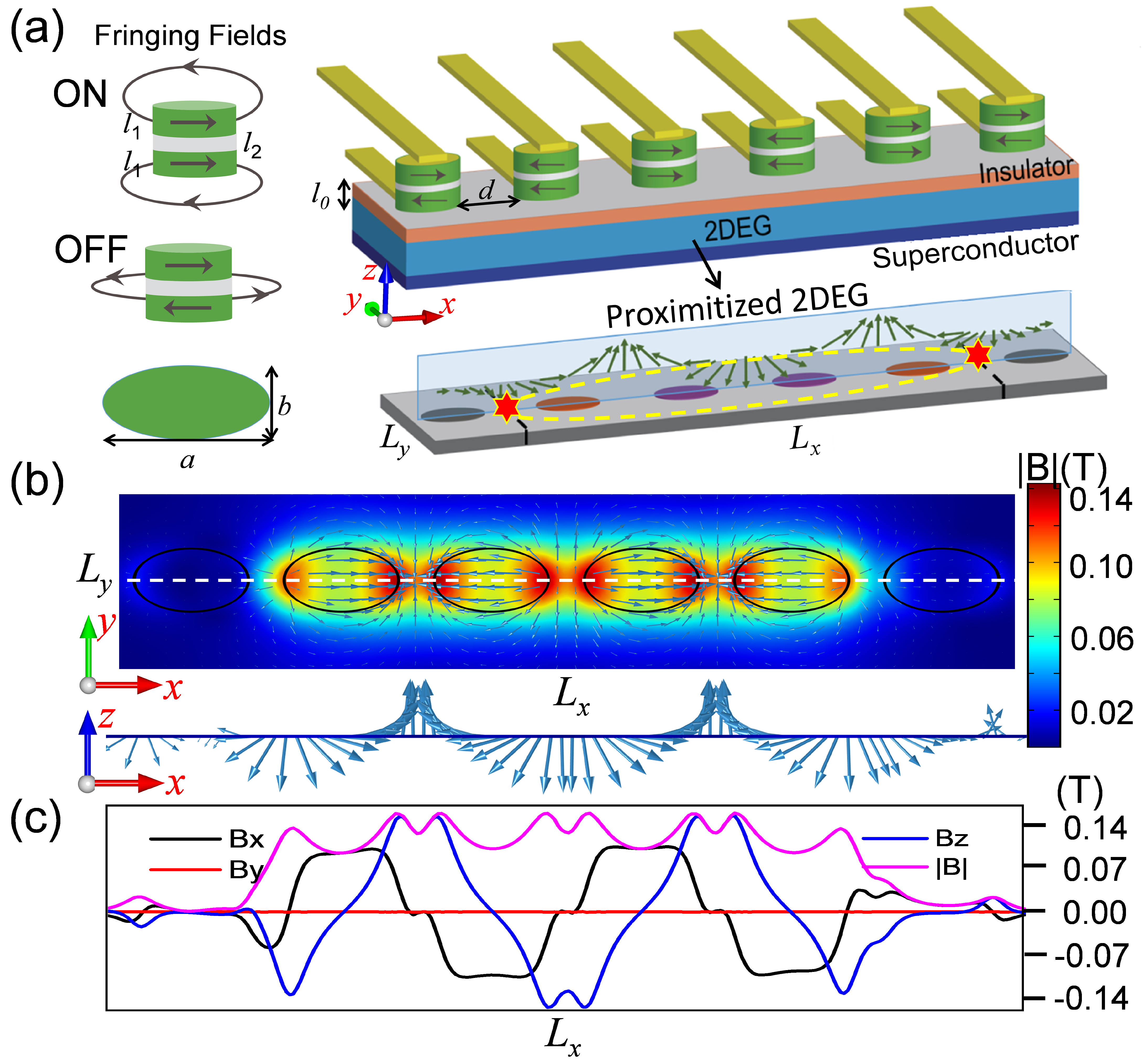}
\caption{(a) Schematic setup. 2DEG is formed next to
the surface of an $s$-wave superconductor. 
Magnetic nanopillars (MNPs) produce magnetic textures, 
which can be tuned by passing currents through the MNP golden contacts and 
switching individual MNPs to the ON or OFF  configuration. 
Each MNP has two magnetic and a nonmagnetic layer, of thickness $l_{1}$
and $l_{2}$, with an elliptical cross section $a\times b$. 
The MNP-2DEG distance is $l_{0}$. Effective topological wires (dashed lines) form in the proximitized 2DEG along the helical-like 
magnetic texture to create Majorana bound states (MBS) (red stars) 
at their ends.  
Switching the MNPs configurations reconfigures the topological wires to transport MBS. (b) The top view (upper) of the simulated fringing field for the structure with $a$ = 160 nm, $b$ = 120 nm, $l_1$ = 7 nm, $l_2$ = 2 nm, 
$d=50$ nm, $l_0$ = 25 nm, $L_x$ = 1500 nm, $L_y$ = 400 nm, and the side view (lower) 
for the magnetic field along the white line in the upper figure. (c) The fringing field along the same
line in (b).
}\label{fig:Scheme}
\end{figure}

However, despite the decades-long use of spin valves in spintronics for magnetically storing and sensing information~\cite{Zutic2004:RMP}, as well as advances in superconducting spin valves~\cite{Buzdin2005:RMP,Linder2015:NP,Eschrig2015:RPP,Eschrig:2019,Alidoust2018:PRB},
there are no prior studies to accurately model how the resulting magnetic textures and fringing fields in realistic structures 
would create and control MBS. 
Here we establish such micromagnetic modeling to examine not just the feasibility of MBS in 
proximity-modified 2DEG, but also to suggest alternative methods to realize braiding in other platforms, including 
those relying on MBS formed in vortices~\cite{Ivanov2001:PRL,Fu2008:PRL}. 
The need for such modeling is 
further motivated by various realizations of nanoscale magnetic textures in the MBS studies:
atomic chains~\cite{Nadj-Perge2014:S,Pawlak2016:NPJQI,Nakosai2013:PRB,Kim2018:SA}, nanomagnets~\cite{Klinovaja2012:PRL,Kjaergaard2012:PRB,Virtanen2018:PRB}, 
domain walls~\cite{Kim2015:PRB,Marra2017:PRB}, skyrmions~\cite{Yang2016:PRB,Gungordu2018:PRB},
or magnetic tips~\cite{Sun2016:PRL,Xu2015:PRL}.

We focus on an array of spin valves in a magnetic nanopillar (MNP) geometry with the underlying magnetic textures controlled by  
spin-transfer torque (STT),
also implemented in commercial magnetic random access memories~\cite{Kent2015:NN,Tsymbal:2012}. 
By passing a current through each MNP, as shown in Fig.~\ref{fig:Scheme}(a), the relative orientation 
of the two magnetic regions can be changed from parallel (ON) to antiparallel (OFF) and thus controlling the resulting fringing fields. 
While the resulting fringing fields, inherent to magnetic arrays, are often ignored, they play a crucial role in 
forming effective topological wires in the neighboring 2DEG (dashed 
lines) with MBS at their ends. In addition to generating Zeeman splitting and particle 
confinement, overcoming the need for a complex network of physical wires, these fringing fields result in synthetic spin-orbit coupling (SOC) in 
the 2DEG~\cite{Fatin2016:PRL,Matos-Abiague2017:SSC}.  Within this scheme, the MBS manipulation relies on STT-controlled magnetic textures, 
without the need for additional contacts and their corresponding risk of  quasiparticle poisoning~\cite{Aasen2016:PRX}. 

We consider CoFe, commonly used in STT and magnetic tunnel junctions~\cite{Tsymbal:2012},  as the magnetic layers in MNPs, with a large saturation magnetization, 
$M_s=1.7\times10^{6}$ A/m, to support strong and controllable magnetic textures. 
CoFe layers, shaped as $160$ nm $\times$ $120$ nm $\times$  $7$ nm elliptical cylinders, 
are separated by a $2$ nm-thick nonmagnetic layer, while each MNPs in an array are spaced $50$ nm apart. The resulting fringing fields in a 2DEG, at $25$ nm below 
MNPs, are simulated using the finite-element method in COMSOL~\cite{COMSOL}, and given in Fig.~\ref{fig:Scheme}(b).
Remarkably, these fringing fields obtained from common ferromagnets 
have a helical-like structure [Fig.~1(b), lower panel] 
similar to that expected to support 
MBS 
in 1D systems~\cite{Lutchyn2010:PRL,Oreg2010:PRL}, even in the absence of a native SOC 
(relevant in some 2DEGs and superconductors~\cite{Zutic2004:RMP,Schnyder2013:PRL}).
The detailed spatial dependence of the fringing fields is shown in Fig.~1(c). 
Unlike common superconducting spin valves~\cite{Buzdin2005:RMP,Linder2015:NP,Eschrig2015:RPP,Eschrig:2019,Alidoust2018:PRB}, our MNPs contain no superconducting 
element. MNPs are located on the top of the structure, separated by $l_0$-thick (25 nm) insulator from the proximitized 2DEG. Therefore, except from the fringing fields, our MNP spin valves do not influence the 2DEG, nor the superconductor below.  

Proximity-induced superconductivity in the 2DEG, which is modified by magnetic textures of an MNP array,  is described by the Bogoliubov-de Gennes (BdG) Hamiltonian,
\begin{equation}
H = \left(\mathbf{p}^2/2m^\ast -\mu \right)\tau_z + \Delta \tau_x + \mathbf{J\left(r\right)}
\cdot\boldsymbol{\sigma} \;,
\label{eq:BdG}
\end{equation}
where $\tau_i$ ($\sigma_i$) are the Nambu (Pauli) matrices in particle-hole (spin) space, and $\mathbf{p}$ and $m^\ast$ are, the momentum and effective mass of the carriers, respectively. The chemical potential, $\mu$~\cite{chemical}, and the proximity induced superconducting gap, $\Delta$, are assumed to be constant. The last term corresponds
to the Zeeman interaction $\mathbf{J(r)}$ = $g^*\mu_B\mathbf{B}$/2, where $g^*$ is the effective $g$-factor, $\mu_{\rm B}$ the Bohr magneton, and $\mathbf{B}$ denotes 
the inhomogeneous magnetic fields generated by the MNP array, 
obtained from micromagnetic modeling of fringing fields. 
To realize MBS with effective spinless pairing, in proximity-induced superconductivity from an $s$-wave superconductor SOC, is required. By performing local spin rotations aligning the spin quantization axis to the local $\mathbf{B}$ direction, Zeeman interaction from Eq.~(\ref{eq:BdG}) is diagonalized $|\mathbf{J\left(r\right)}| \sigma_z$ and accompanied by a non-Abelian field that yields synthetic SOC~\cite{Korenman1977:PRB,Tatara1997:PRL,Braunecker2010:PRB,Pientka2013:PRB,Klinovaja2013:PRL}.

Tunable magnetic textures in  Eq.~(\ref{eq:BdG}) generalize the common MBS implementation in 1D 
semiconductor nanowires  with a homogeneous $\mathbf{B}$-field and the resulting condition for a topological phase transition, 
$E_\mathrm{Zeeman}=(\mu^2+\Delta^2)^{1/2}$~\cite{Lutchyn2010:PRL,Oreg2010:PRL}. In our case, the formation of topological regions 
is approximately determined by~\cite{Fatin2016:PRL,Matos-Abiague2017:SSC},
\begin{eqnarray}
\label{eq:top}
\left|\mathbf{J\left(r\right)}\right|^2 = \left[\mu - \eta\left(r\right)\right]^2 + \Delta^2, \\
\eta\left(r\right) = \frac{\hbar^2}{8 m^\ast \left|\mathbf{J\left(r\right)}\right|^2} \sum_{i=1}^2\frac{\partial \mathbf{J\left(r\right)}}{\partial x_i}\cdot \frac{\partial \mathbf{J\left(r\right)}}{\partial x_i},
\label{eq:eta}
\end{eqnarray}
where $\eta$ represents an effective shift in the chemical potential due to local changes of the magnetic texture. 
For a
homogeneous $\mathbf{B}$-field, $\eta \rightarrow$ 0, this generalized topological condition reduces to the previous one
determining the topological 
transition in quantum wires and rings~\cite{Lutchyn2010:PRL,Oreg2010:PRL,Scharf2015:PRB,Alicea2012:RPP}. If we rewrite Eq.~(\ref{eq:top}) as 
$\mathrm{P}=\left|\mathbf{J\left(r\right)}\right|^2 - (\left[\mu - \eta\left(r\right)\right]^2 + \Delta^2)$, the set of positions P=0, where the topological condition 
is fulfilled, forms a contour that 
separates topological (P $>0$) and trivial (P $<0$) domains,
as shown 
in Fig.~2 and discussed below. According to the bulk-boundary correspondence, 
localized states emerge at the border between the topological  and trivial domains. 
Depending on the specific geometry of the closed contour, the edge states can eventually collapse
into MBS localized at the ends of the topological contour when it approaches the quasi 1D limit. 

\begin{figure}[t]
\centering
\includegraphics*[width=0.5\textwidth]{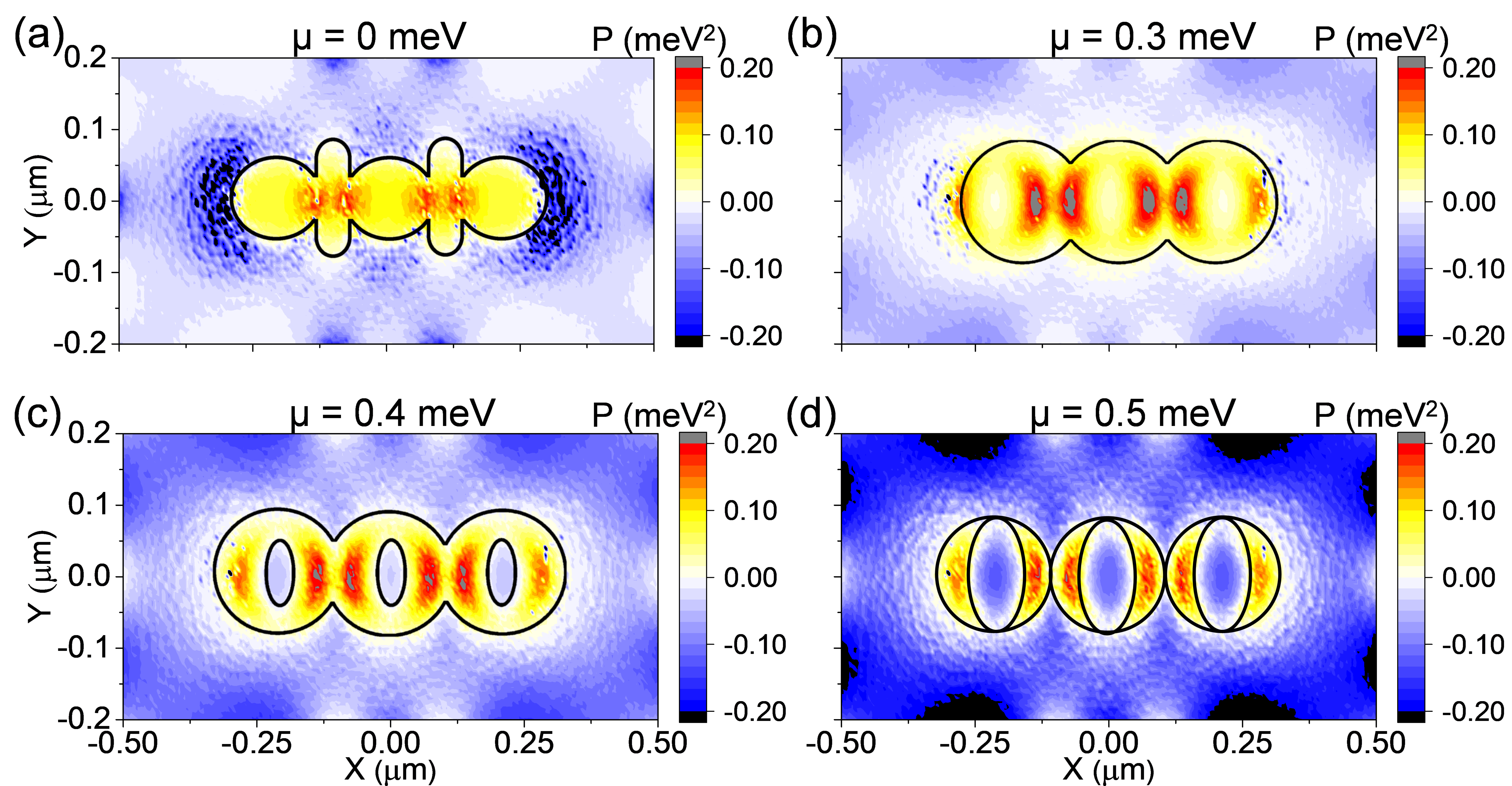}
\caption{Evolution of topological contours with a chemical potential for the 3-ON MNPs: (a) $\mu$ = 0, (b) $\mu$ = 0.3 meV, (c) $\mu$ = 0.4 meV, and (d) $\mu$ = 0.5 meV. The black lines in (a) - (d) indicate P=0, giving the effective topological wires. All the MNP parameters are the same as those in Fig. 1(b),
$\Delta$ = 0.1 meV, $m^*$=0.026 $m_e$, and $g^*$ = 120.}
\label{fig:Contours}
\end{figure}

For the MBS formation and manipulation, 
our MNP/2DEG/ superconductor platform must be carefully designed. 
A suitable choice is given by InAs/Al-based 2DEG/superconductor systems. Recent experiments show a robust proximity-induced 2D superconductivity 
in InAs that forms transparent contacts with Al and yields a very large critical current in 
Josephson junctions~\cite{Shabani2016:PRB,Mayer2018:P}. With 
all-epitaxial growth, proximity-induced superconducting gap, $\Delta$, attains nearly the bulk value of Al ($\Delta_\mathrm{Al}\sim0.2$ meV)~\cite{Mayer2018:P}. 
To strengthen the influence of tunable magnetic textures on 2DEG superconductivity and allow for a larger MNP-2DEG separation ($l_0$, see Fig.~\ref{fig:Scheme}), 
$g^*$-factor in InAs can be enhanced by doping (as well as due to the orbital effects~\cite{Winkler2017:PRL}. 
In $n$-doped (In,Mn)As,  $g^*>120$ was realized~\cite{Zutic2004:RMP,Zudov2002:PRB}, 
while InAs$_{1-x}$Sb$_x$ family, with interesting
topological properties~\cite{Winkler2017:PRL}, attains $g^*\sim140$ for $x\sim 0.6$~\cite{Svensson2012:PRB}. To describe our  2DEG, 
in BdG Hamiltonian we choose $g^*$ = 120, $m^*$ = 0.026 $m_0$ ($m_0$ is the bare electron mass), and $\Delta=0.1$ meV.
Even a much larger $g^*\sim 300$ was demonstrated in (Cd,Mn)Te 2DEG, supporting a strong influence of the fringing fields~\cite{Betthausen2012:S},
but transparent superconducting junctions and robust proximity-induced superconductivity have not yet been demonstrated, 
thus making the InAs-based 2DEG systems more promising for MBS.   

To obtain MBS in the MNP/2DEG/Al platform, the structure parameters ($a$, $b$, $d$, $l_0$, $l_1$, $l_2$) of the MNP array need to be tuned 
to generate the appropriate magnetic texture and drive the system into the topological regime. Given a large parameter space, this is a difficult task 
from the rigorous analysis in which the micromagnetic modeling of the fringing fields is used as an input for $\mathbf{B}$ of Eq.~(\ref{eq:BdG}). 
However, this procedure is considerably simplified without solving the BdG equation and instead examining the generalized topological 
condition in Eq.~(\ref{eq:top}).
Furthermore, we can reduce the system size to first optimize these structure parameters based on 3 ON-MNPs.  

After some tests, we find that fringing fields with $a$ = 160 nm, $b$ = 120 nm, $d$ = 50 nm, $l_0$ = 25 nm, $l_1$ = 6 nm, and $l_2$ = 2 nm can induce 
effective topological wires for a large range of $\mu$. Figure~\ref{fig:Contours} shows the resulting evolution of the topological contours with a chemical potential. 
The black P=0 contours indicate the boundary between the trivial (outside) and nontrivial (inside) regions, giving the effective wires. 
With $\mu$ from 0.0 meV to 0.3 meV, there is a single 
continuous effective wire where at its two ends MBS are expected to emerge. When $\mu$ is increased to 0.4 meV, 
small topologically trivial regions appear inside the outer contour, 
whose geometry then becomes unfavorable for the formation of MBS.
When $\mu$ is up to 0.5 meV, there are no long continuous topological contours, indicating the absence of MBS. 

\begin{figure}[t]
\centering
\includegraphics*[width=0.48\textwidth]{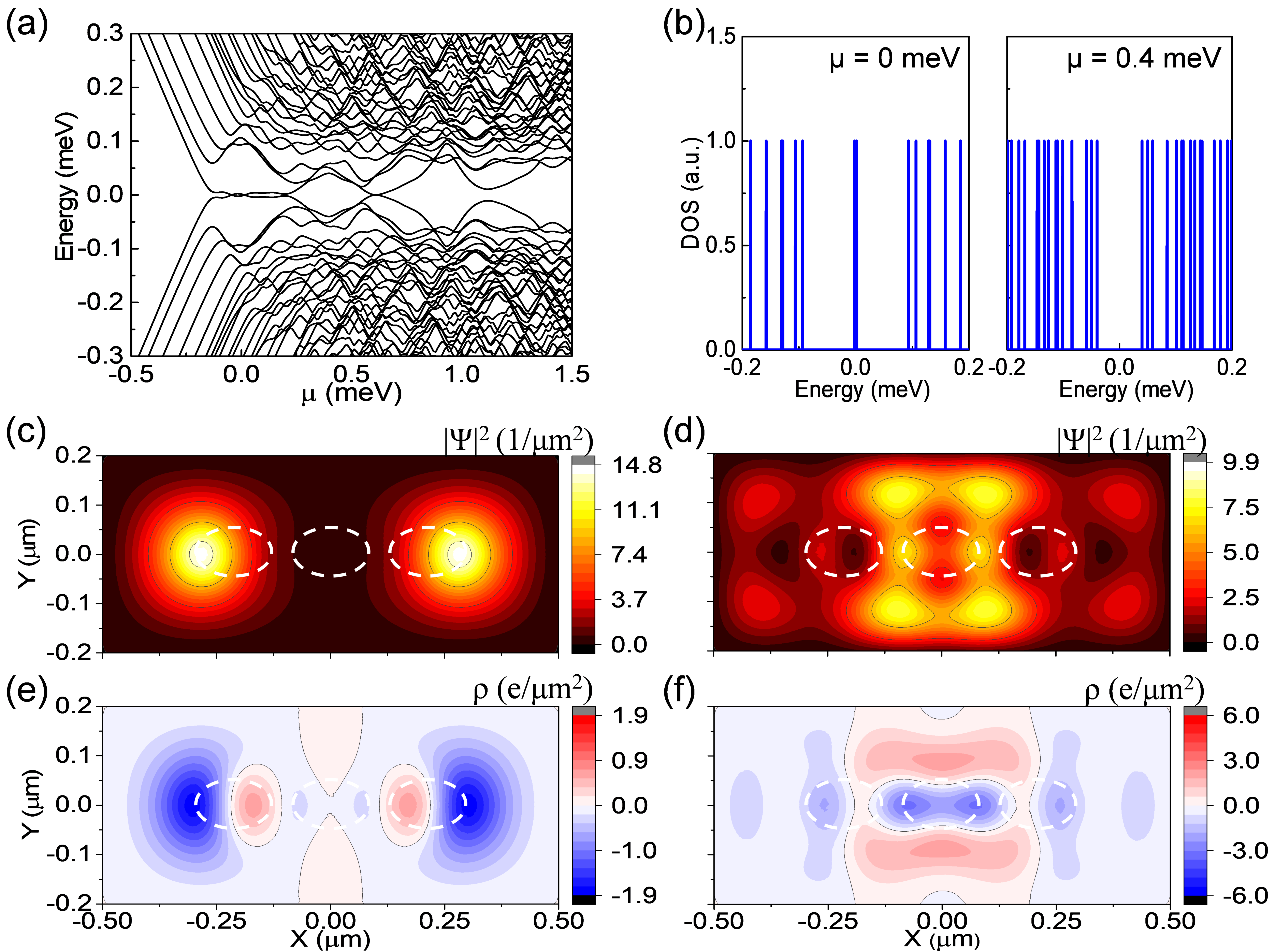}
\caption{(a) Low-energy spectrum as a function of the chemical potential 
for a system in Fig. 1 with 3-ON MNPs. (b) The density of states from (a). 
(c) and (d) The probability density 
for the lowest energy states with $\mu$ = 0 (topological) and $\mu$ = 0.4 meV (trivial), respectively. 
The black lines in (c) and (d) indicate the contours with the values in the color bars. (e) and (f) charge densities for the lowest energy states 
with $\mu$ = 0  (topological) and $\mu$ = 0.4 meV (trivial), respectively. Dashed lines in (c)-(f): the MNP array, black lines in (e) and (f): zero contour values. 
The parameters are taken from Fig.~2.} 
\label{fig:3MNP}
\end{figure}

Recognizing that topological contours can provide a computationally-efficient guidance for MBS, but do not necessarily give the exact parameters
for their existence, we turn to the solution of the BdG equations for the MNP/2DEG/Al system. 
The complexity of the simulated magnetic texture only permits a numerical determination of MBS existence. We solve an eigenvalue problem for the BdG Hamiltonian from Eq.~(\ref{eq:BdG}) using a fourth order finite-difference method~\cite{Fatin2016:PRL}.  
The resulting low-energy spectrum corresponding to $\mu$ 
for the 3-ON MNP/2DEG/Al system is shown in Fig.~\ref{fig:3MNP}(a), where
 $\mu = 0$  indicates the bottom of the conduction band of the  
 2DEG by itself (without any proximity effects).

Nearly zero energy states (ZES) are not necessarily MBS. For example, Andreev bounds states can also occur at zero energy~\cite{Liu2017:PRB,%
Eschrig2015:RPP,Eschrig:2019,Alidoust2018:PRB,Crepin2014:PRB,Kashiwaya2000:RPP,Hu1994:PRL,Chen2001:PRB,Zutic2000:PRB,Alidoust2015:PRB}. 
Here ZES denote any states with (nearly) zero energy, being MBS or not and are related to the well-known zero bias 
conductance peak~\cite{Sengupta2001:PRB,Law2009:PRL}. ZES emerge in the superconducting gap for $\mu$ within the interval from $-0.12$ to $0.24$ meV. 
This is in good agreement with the range of chemical potentials in which the shape and dimensions of the topologically nontrivial zones surrounded 
by the computed topological contours favor the formation of MBS. According to Eq.~(\ref{eq:top}) [see also Fig.~2], 
such a region extends up to $\mu\approx 0.3$ meV, a value slightly higher than the value $0.24$ meV observed in the spectrum. 
This corroborates the usefulness of Eq.~(\ref{eq:top}) for determining the region of system parameters supporting the formation of MBS.

A closer look at the low-energy BdG spectrum in Fig.~3(a) reveals oscillations in the splitting of the approximate ZES as a function $\mu$. 
A similar behavior is well-known for MBS in semiconductor nanowires
~\cite{Lim2012:PRB,Prada2012:PRB,DasSarma2012:PRB,Rainis2013:PRB}. 
For a finite wire, the wavefunctions of the two MBS localized near its ends overlap and hybridize leading to the finite energy 
that decays exponentially with the increasing length of the nanowire and oscillates with the changes in chemical potential. 
This is further shown in  Supplemental Material (see Ref.~\cite{SM18}).

In addition to ZES, the existence of MBS can also be supported by the spatially localized probability density, $\left|\Psi\right|^2$ = $\left|u\right|^2$ + $\left|v\right|^2$, 
and a neutral charge density, $\rho$ = $\left|u\right|^2$ - $\left|v\right|^2$, where $u$ and $v$ are particle and hole components of its 
wavefunction, respectively~\cite{Scharf2015:PRB,Alicea2012:RPP,Ben-Shach2015:PRB}. 
For comparison, we calculate $\left|\Psi\right|^2$ and $\rho$ of both topological ($\mu$ = 0) and trivial states ($\mu$ = 0.4 meV).
While ZES are clearly present for $\mu$ = 0, no such states exist in the gap at $\mu$ = 0.4 meV in Fig.~3(b). 
Among the lowest energy states in Figs.~3(c) and (d), 
there is also an obvious difference between their spatial distribution in the topological regime ($\mu=0$), 
where the $\left|\Psi\right|^2$  is localized at the two ends of the effective wires, and in the trivial regime ($\mu=0.4$ meV), where the 
$\left|\Psi\right|^2$ is effectively spread over the whole system. 
The MBS charge density 
(at $\mu$ = 0) in Fig.~3(e) is much smaller over the entire structure than for the trivial states ($\mu$ = 0.4 meV) in Fig.~3(f). 

\begin{figure}[t]
\centering
\includegraphics*[width=0.48\textwidth]{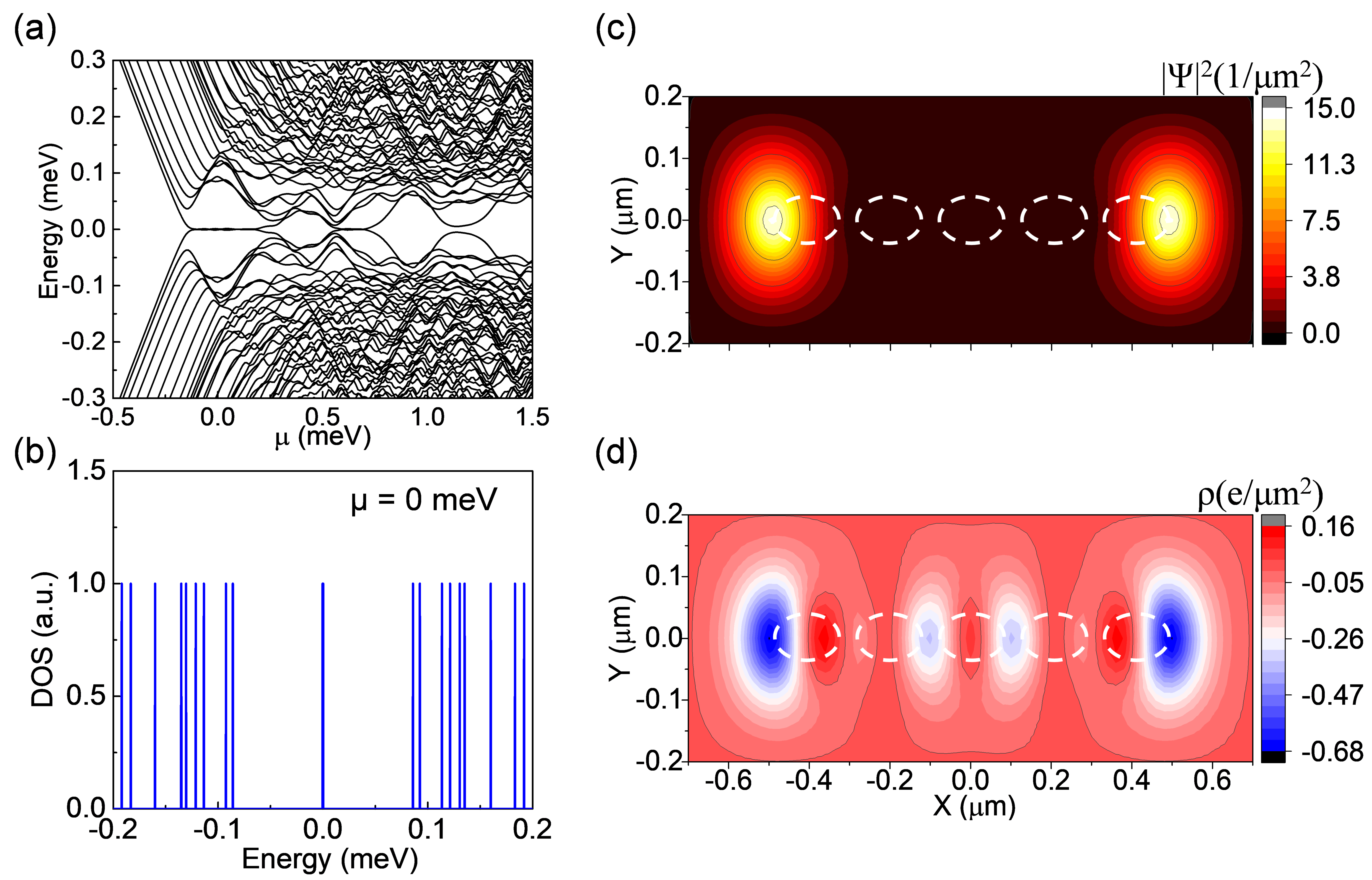}
\caption{(a) Low-energy spectrum as a function of the chemical potential 
for a system in Fig. 1 with 5-ON MNPs. (b) The density of states from (a). 
(c)  and (d) The probability density and charge density for the lowest energy states with $\mu$ = 0. 
The black lines in (c) indicate the contours with the values in the color bars, 
and in (d) the zero-contour values. Dashed lines in (c) and (d) denote the MNP array.
The parameters are taken from Fig.~2.} 
\label{fig:5MNP}
\end{figure}

In our case, the length of the effective wire for 3 MNPs is about 540 nm, which is  too short to separate well the two MBS. 
However, as in the semiconductor nanowires, we show in  Ref.~\cite{SM18}) that the oscillations of the ZES 
splitting are strongly suppressed in longer effective topological wires with the increased number (5, 6, 7, 9) of MNPs.
As expected,  with the increase in the system size, the MBS become more localized at the two ends of the effective wire~\cite{Alicea2012:RPP}.
This can already bee seen in Fig.~\ref{fig:5MNP} for 5-ON MNPs, showing also that in the topological regime the MBS charge density 
is reduced as the system size is increased, approaching the $\rho \rightarrow 0$ limit, as expected for  MBS in an infinitely long wire. 
Based on the above results for the ZES, localized probability density, and charge neutrality, there is comprehensive support for the MBS formation in 
MNP/2DEG/Al systems. 

Through magnetic textures and the emergent synthetic SOC all the ingredients required for MBS are realized. However, native SOC is also inherent to InAs-based 2DEG and
may influence the MBS formation. To assess such SOC effect, we consider a typical value for the Dresselhaus SOC~\cite{Zutic2004:RMP,Fabian2007:APS} with strength of $\gamma=40$ meV\AA~\cite{Fatin2016:PRL},
but the MBS formation remains largely unchanged, as shown in Ref.~\cite{SM18}.  

\begin{figure}[t]
\centering
\includegraphics[width=0.48\textwidth]{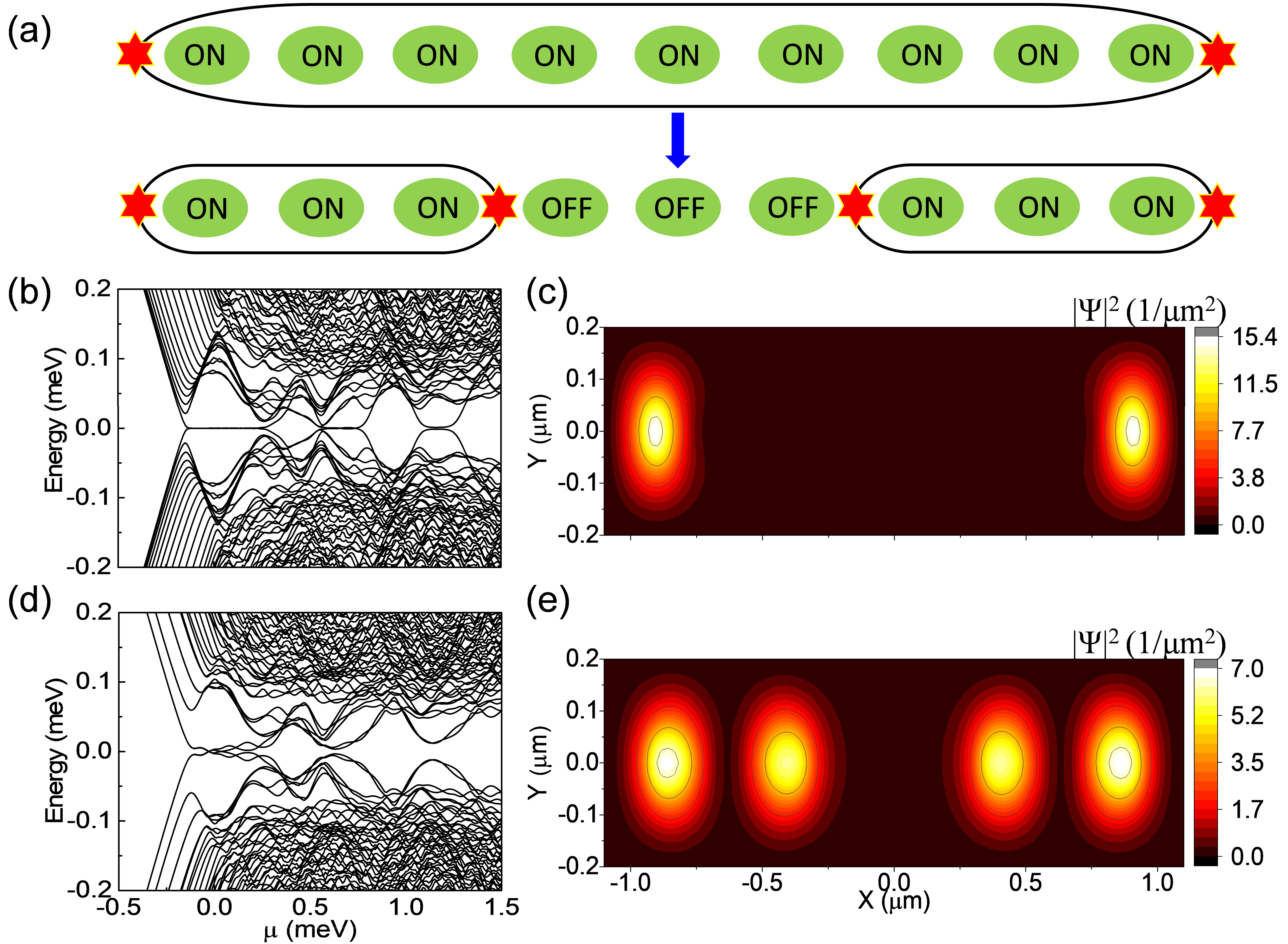}
\caption{(a) Schematic of the MBS control. In 9-ON MNPs, MBS (red stars) are localized at the ends of the effective wire (black lines). 
By switching the middle 3 MNPs from ON to OFF,  the effective wire is  broken and two additional MBS emerge. 
(b) Low-energy spectrum for a system in Fig.~1 with the upper MNPs array in (a). 
(c) The probability density for the lowest energy states in (b), $\mu = -0.1$ meV. 
(d) Analogous to (b), but with the lower MNPs array in (a). 
(e) The probability density for the lowest energy states in (d), $\mu = -0.1$ meV. 
The parameters are taken from Fig. 2.}
\label{fig:Fusion}
\end{figure}

Another interesting phenomenon in MNP/2DEG/Al systems is the reentrant topological regime with an increase in $\mu$. In Fig.~3(a) this appears near $\mu=0.6$ meV.
The calculated probability and the charge densities 
confirm that ZES at $\mu \approx 0.6$ meV are indeed MBS~\cite{SM18}.
The origin of these MBS is related to the presence of multiple subbands and, therefore, cannot be explained by the  
approximate topological condition in Eq.~(\ref{eq:top}), 
derived  within the single-subband approximation. 

To examine this presence of the reentrant topological regime, we 
consider longer effective wires than from previous 
3-ON MNP arrays. This can be clearly seen in Figs.~\ref{fig:Fusion}(a)-(c) for 9-ON MNPs and  $\approx 2$ $\mu$m long effective wires.
With a smaller overlap of the two MBS, the ZES range is enhanced and now even shows the third reentrant regime for $\mu >1$ meV, while the oscillations in the
ZES splitting are visibly reduced. 

\begin{figure}[b]
\centering
\includegraphics[width=0.48\textwidth]{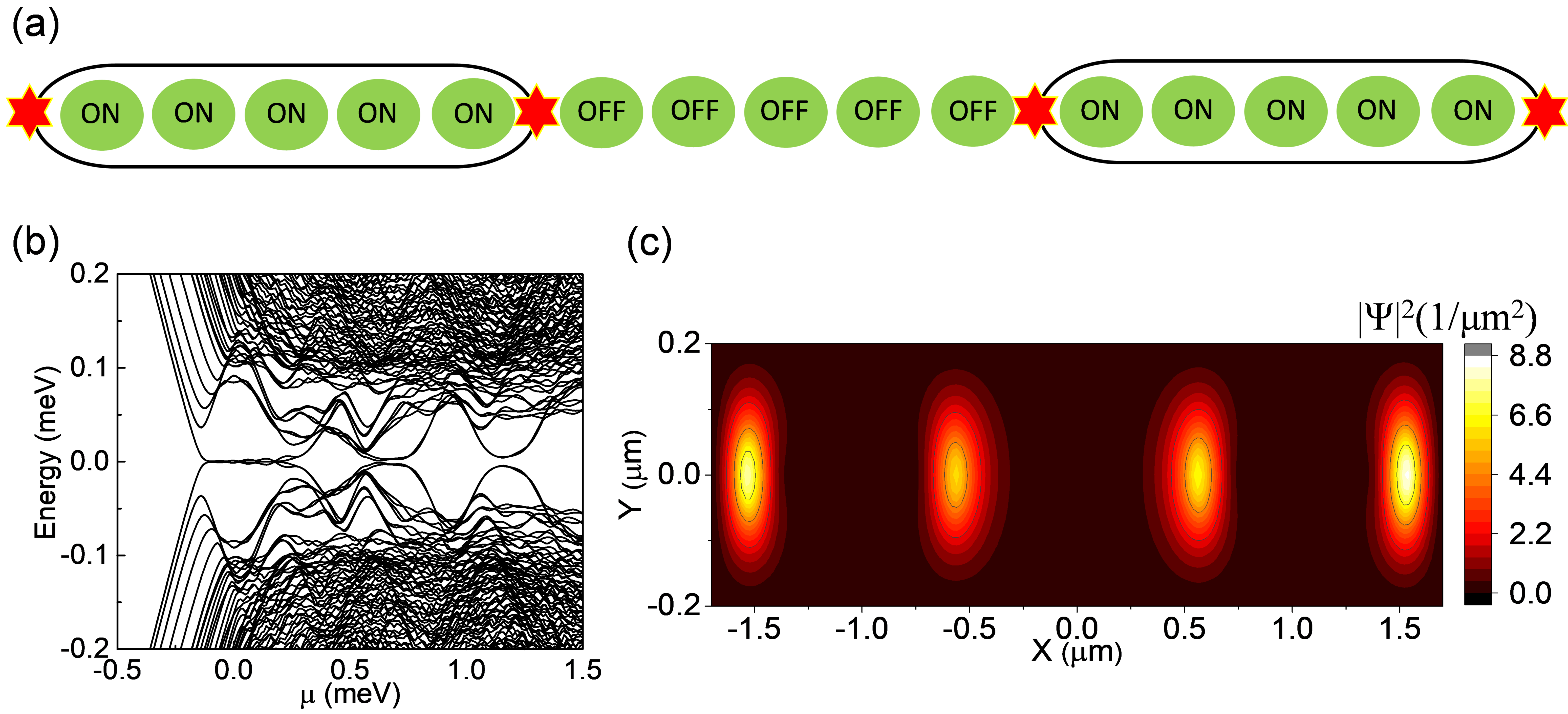}
\caption{(a) Schematic of the two separated effective wires by controlling the MNPs based on 15 MNP array. 
MBS (red stars) emerge at the two ends of each effective wire. 
(b) Low-energy spectrum for a system in Fig.~1 but with the 15 MNP array shown in (a). 
(c) The probability density for the lowest energy states in (b) with $\mu = 0$. 
The parameters are taken from Fig. 2.}
\label{fig:15Fusion}
\end{figure}

However, the tunability of magnetic textures to reconfigure these longer effective wires provides more than just an opportunity
for improved MBS signatures. Instead, as schematically shown in Fig.~\ref{fig:Fusion}(a), these wires could be used to test the
non-Abelian statistics through fusion rules of MBS~\cite{Aasen2016:PRX}. By STT switching the middle 3 MNPs from ON  to OFF states, 
the effective wire breaks into two shorter and disjoint ones, as indicated by the effective topological contours. With this transformation,  
two energy states exist in the band gap [Fig.~\ref{fig:Fusion}(d)],
and one MBS pair emerges at the ends of each effective wire,
a signature as shown in Fig.~\ref{fig:Fusion}(e). Because these two effective wires are short and close together, the overlap between MBSs is large, 
resulting in visible oscillations in ZES splitting in Fig.~\ref{fig:Fusion}(d). 

For even longer wires with 15 MNPs  we expect improved MBS signatures. Indeed, our results from Fig.~\ref{fig:15Fusion} confirm 
that these oscillations are strongly suppressed and the MBS pairs at the end of the two longer effective wires are better localized. 
With the reversible control of MNP arrays,  
another STT switching of the middle  3 MNPs from OFF  to ON states  in the 9-MNP configuration from Fig.~\ref{fig:Fusion}
returns to the lower scheme in Fig.~\ref{fig:Fusion}(a) to its initial configuration, causing the fusion of the additional MBS pair while recovering the initial 
9-ON MNP effective wire with two end MBS. For wires with 15 MNPs from Fig.~\ref{fig:15Fusion}, an analogous fusion of the MBS pair
would require switching the middle  5 MNPs from OFF  to ON states. 

\begin{figure}[t]
\centering
\includegraphics[width=0.48\textwidth]{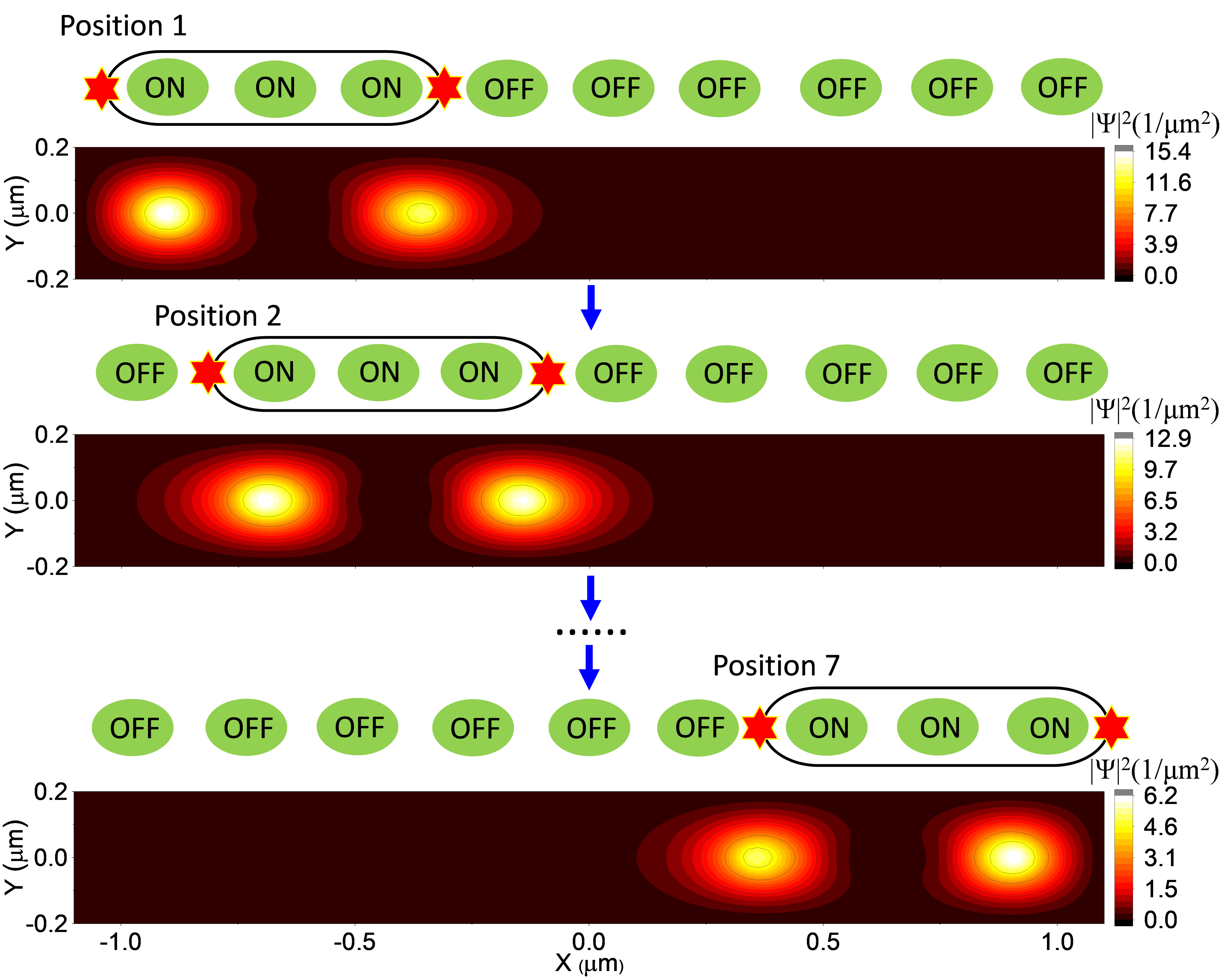}
\caption{Moving MBS by controlling the MNPs based on 9 MNP array. MBS (red stars) can be moved from the left-most part (Position 1) to the right-most part (Position 7) 
by switching MNPs. The parameters are taken from Fig.~2}
\label{fig:Move}
\end{figure}

By MNP switching one can move and manipulate both the position and overlap of the different MBS 
and, therefore, enable the implementation of fusion and, eventually, braiding. While the full potential 
of this approach benefits from the scalability of the 2D MNP arrays, a transfer of MBS across an effective
topological wire can be already realized in a simple 1D array as illustrated in Fig.~\ref{fig:Move}, suggesting
the feasibility of future generalization for braiding and fusion of MBS as a test of their non-Abelian statistics. 

Our proposal for realizing 
reconfigurable topological wires closely relies on the recent advances in spintronics to control magnetic textures. Since 2D arrays of similar MNPs, individually addressed by STT to change between their ON and OFF states are already commercially available~\cite{Tsymbal:2012}, it is also possible to envision how our platform would enable MBS 
braiding~\cite{Fatin2016:PRL,Matos-Abiague2017:SSC} compatible with available materials and device processing. Our framework, which combines accurate modeling of the magnetic textures as an input for the solution of the BdG equations, is also flexible enough to investigate other MBS platforms with magnetic elements or even consider manipulating MBS confined in vortices. Rather than using STT, it may also be possible to control magnetic proximity effects in arrays of magnets by gating~\cite{Lazic2016:PRB,Zutic2018:MT}.

With the experimental progress aimed at confirming our predictions, an important question to address pertains to various forms of the disorder and their effect on the robustness of the MBS formation and control. Some general intuition and encouraging trends are already available by contrasting the role of disorder in the topological regime with the better studied trivial regime, since the disorder may even promote the MBS formation~\cite{Adagideli2014:PRB,Habibi2018:PRB}. Our preliminary studies corroborate that by showing that the disorder in the orientation of the magnetization in the two regions of a spin valve~\cite{Fatin2016:PRL}, or in the size of MNPs 
has only a relatively weak effect on MBS. Considering superconducting systems and proximitized 2DEGs or topological insulators~\cite{Schuffelgen2017:P}, 
where the fringing fields can play an important role, provides an interesting opportunity to revisit the self-consistent 
description~\cite{Valls2010:PRB,Halterman2018:PRB,Zyuzin2016:PRB,Banerjee2018:PRB} and include changes to the Meissner regime due to finite and nonuniform magnetic regions~\cite{Zutic1997:JCP,Halterman2001:PRB}.
While our focus was on systems with large effective $g$-factors, this is not the fundamental limitation. With a different design of magnetic arrays and their closer distance to the region with a
proximity-induced superconductivity the resulting fringing fields can exceed 1 tesla and also support MBS manipulation in materials with much smaller $g$-factors.

We thank J. Shabani and A. K. Kent for valuable discussion about their related experimental efforts. We also thank W. C. Yu for his valuable discussion about the micromagnetic simulations.
This work is supported by DARPA Grant No.  DP18AP900007 (T. Z., N. M., J. E. H., A. M.-A., and I. \v{Z}.), 
US ONR  Grant No. N000141712793 (A. M.-A and I. \v{Z}.), and the UB Center for Computational Research.
I. \v{Z}. acknowledges the hospitality of the Institute for Theoretical Physics, University of Regensburg. 

\bibliography{MBS_MNP_resubmit}

\begin{thebibliography}{85}
\bibitem [{che()}]{chemical}%
  \BibitemOpen
  \href@noop {} {}\bibinfo {note} {More accurately, $\mu$ is the energy
  difference from the band bottom at B=0 to the chemical potential, chosen to
  be at zero energy.}\BibitemShut {Stop}%
\end{thebibliography}

\end{document}